\documentclass[journal]{IEEEtran}

\ifCLASSINFOpdf
   \usepackage[pdftex]{graphicx}
  
  \graphicspath{{../pdf/}{../jpeg/}}
  
   \DeclareGraphicsExtensions{.pdf,.jpeg,.png}
\else
  
   \usepackage[dvips]{graphicx}
  
   \graphicspath{{../eps/}}
 
   \DeclareGraphicsExtensions{.eps}
\fi

\usepackage{amsmath}
\usepackage{cite}
\usepackage{diagbox}
\begin{document}

\title{Cluster-Based Cell-Free Massive MIMO Systems: A Novel Framework to Enhance Spectral Efficiency with Low Complexity}

\author{Reza~Roshanghias,
        Reza~Saadat 
        
\thanks{}
\thanks{}
\thanks{}}

\markboth{}%
{Shell \MakeLowercase{\textit{et al.}}: Bare Demo of IEEEtran.cls for IEEE Journals}

\maketitle

\begin{abstract}
The issue of diminished spectral efficiency (SE) of the downlink (DL) transmission in distributed cell-free massive MIMO (CF-mMIMO) systems poses a significant challenge in terms of user equipment (UE) performance when compared to their centralized CF-mMIMO  counterparts. The primary root cause of this issue can be attributed to the reduced efficacy of distributed precoders, which are devised using local channel state information (CSI) in distributed systems. This reduced efficacy becomes particularly pronounced in terms of interference mitigation when compared to centralized precoders. To address this issue, this paper proposes a novel architectural framework for CF-mMIMO systems, referred to herein as the "cluster-based structure." Within this innovative structure, a hybrid amalgamation of centralized and distributed configurations is employed, complemented by the introduction of a unique cluster arrangement for the access points (APs) within the network. In this design, the CSI of APs within each cluster is collectively shared within a local processor unit. Consequently, by harnessing this enhanced repository of local channel information, local precoders are formulated, which facilitate more effective interference mitigation with reduced computational complexity compared to the centralized approach. This approach ultimately results in a significantly augmented SE when contrasted with the distributed architecture. In this paper, the DL signal-to-interference-plus-noise ratio (SINR) of the UEs in this new architecture is derived analytically, and two precoders, namely maximum ratio (MR) and minimum mean square error (MMSE), are proposed for the new architecture. Furthermore, an elucidation of the computational complexity associated with the MMSE precoder in the context of the cluster-based framework will be presented, drawing a comparative analysis with both centralized and distributed structural configurations. The simulation results unequivocally demonstrate that within the cluster-based framework, the optimal SE for the network is attained when utilizing four clusters in conjunction with the MMSE precoding technique, leading to a notable reduction in computational complexity exceeding $85\%$. Importantly, this approach surpasses the SE performance of the centralized structure. 
\end{abstract}


\IEEEpeerreviewmaketitle

\section{Introduction}
To support  the significant growth of smart devices and the exponential proliferation of mobile data traffic, the deployment of dense communication networks is being envisaged. One emerging approach for constructing wireless networks that facilitate ubiquitous connectivity is cell-free massive MIMO (CF-mMIMO) systems. These systems involve a large number of collaborative access points (APs) working together to serve users in a wide area, without the presence of cell boundaries \cite{ngo2017cell, bjornson2019making, nayebi2017precoding, zhang2019cell,zhang2020prospective, mai2018pilot}. This approach eliminates the inherent form factor constraints associated with conventional co-located massive MIMO (mMIMO) systems, thereby offering enhanced flexibility. CF-mMIMO represents a more comprehensive concept that combines the advantages of two fundamental technologies: mMIMO, which leverages favorable propagation characteristics, and Network MIMO, which enables more uniform user performance through coherent signal processing among multiple distributed APs. This technology can be implemented within a cloud radio access network (C-RAN) architecture. In comparison to traditional Network MIMO, the key distinguishing feature of CF-mMIMO lies in its operating regime, characterized by a significantly greater number of APs than user equipment (UEs), as well as the utilization of mMIMO-inspired transmission protocols. Recognized as one of the most promising beyond 5G technologies, CF-mMIMO exhibits the ability to suppress multi-user interference and exploit macro diversity, thereby offering a nearly uniform service to UEs \cite{zhang2019cell}, \cite{zhang2020prospective}, \cite{akbar2018location}.

In addition to the considerable advantages provided by CF-mMIMO systems, there are certain inherent challenges in deploying and establishing this system. One primary obstacle relates to the significant computational complexity involved in developing signal-processing algorithms for the implementation of a centralized architecture in the system \cite{buzzi2016survey}. In the centralized architecture of CF-mMIMO systems, signal-processing tasks are performed exclusively in the central processing unit (CPU), while the APs function as relays or radio remote heads (RRH). Within this architecture, the CPU retrieves channel state information (CSI) for all UEs within the network, and utilizes this collective CSI to pre-code the encoded signals for all UEs in the downlink (DL) \cite{demir2021foundations}. As previously stated, the primary drawback of the centralized architecture in CF-mmMIMO systems lies in its substantial computational complexity.

The distributed architecture of CF-mMIMO systems was proposed to reduce the significant complexity of the centralized architecture. In the distributed architecture, each AP in the network utilizes the local processing unit for communication with all UEs in the network. In this architecture, the signals from UEs are encoded in the CPU and pre-coded in APs based on the local CSI of each AP. Thus, unlike the centralized architecture, the APs are involved in pre-coding the UE signals based on their local CSI \cite{demir2021foundations}. While the distributed architecture offers lower computational complexity compared to the centralized architecture, it also results in lower spectral efficiency (SE) for UEs. The primary cause for this phenomenon stems from the distributed architecture, whereby the APs solely utilize the localized CSI. Consequently, the APs construct their precoders based solely on this specific local information, rendering it unfeasible for each AP to effectively mitigate the interference originating from other APs within the network. As a result, the overall SE of UEs diminishes \cite{demir2021foundations}.

Limited research has been conducted thus far to enhance the SE of UEs in distributed CF-mMIMO systems. In the study by the authors \cite{miretti2021team, miretti2022team, miretti2021precoding}, a unique precoder called team precoder is introduced, which leverages the team theory to devise the precoder. Additionally, in papers \cite{gouda2020distributed, atzeni2020distributed} the precoders for UEs are formulated by fostering collaboration among APs.

The main objective of this paper is to improve the SE of user equipment UEs within a distributed CF-mMIMO system. This enhancement will be achieved by employing a novel approach that ensures low computational complexity, as compared to the traditional centralized architecture of the system. To achieve this aim, we propose a cluster-based architecture for a CF-mMIMO system, wherein the precoder for each UE is designed by utilizing the collective local CSI obtained from all APs within each cluster. To establish a cluster-based CF-mMIMO architecture, the initial step involves the creation of a distinctive cluster comprising multiple APs and a local processing unit. These components are interconnected through front-haul links for seamless information exchange. Subsequently, these clusters are interconnected to a central unit using front-haul links to receive encoded signals. Within each cluster, the APs function as relays or RRH, while the local processing unit takes into account the locally acquired CSI from all APs in the cluster to design an appropriate precoder. The primary purpose of these clusters is to cater to all UEs within the network.

The paper makes several significant contributions:

\begin{itemize}
\item  we present an novel framework for CF-mMIMO systems, known as cluster-based CF-mMIMO systems and we outline the key components comprising this framework. Furthermore, an in-depth analysis of the DL transmission within this novel architecture is conducted
\item  we analytically calculates the SE expression for this newly proposed architecture.
\item  we introduces the Minimum Mean Square Error (MMSE) and Maximum Ratio (MR) precoders specifically designed for the novel architecture.
\item  we employ an analytic approach to determine the computational complexity associated with MMSE precoders designed for cluster-based architecture. Furthermore, we undertake a comparative analysis, specifically examining the computational complexity of this MMSE precoder in relation to both centralized and distributed architectures.
\item  we conduct a numerical comparison of the cluster-based architectural framework with two centralized and distributed structural configurations, assessing both computational complexity and spectral efficiency..
\end{itemize}

The subsequent sections of this paper are structured as follows: Section \ref{section two} provides an exposition on DL data transmission in cluster-based Cell-free mMIMO systems, including the presentation of a closed-form expression for downlink SE. Section \ref{section three} investigates MR and MMSE precoders specifically for the cluster-based architecture. The computational complexity associated with MMSE precoders in cluster-based frameworks is examined in Section \ref{section four}. In Section \ref{section five}, we present the numerical results derived from our analysis. Finally, Section \ref{section six} outlines the significant conclusions drawn from this article.

Notations:
In this context, $x$ and ${\bf{x}}$ represent a scalar and vector respectively.
The symbol $(.)^*$ is used to denote the complex conjugate, $(.)^{\rm{T}}$ represents the transpose, and $(.)^{\rm{H}}$ signifies the Hermitian transpose.
The modulus operator is denoted by $\left| x \right|$, $\left\| {\bf{x}} \right\|$ represents the $L_2$ vector norm.
The expected value of its argument is denoted by ${\rm{E}}\left\{ {\left. . \right\}} \right.$, 
and the complex Gaussian distribution for a random variable with zero mean and variance $\sigma^2$ is denoted as $\mathcal{N_{C}}(0,{\sigma^2})$.
Finally, the identity matrix with dimension $N \times N$ is represented by ${\bf{I}}_N$.

\section{System Model} \label{section two}

We consider DL transmission in a cluster-based CF-mMIMO system which consists of $M$ distinct clusters, $L_T$ APs and $K_T$ UEs. The cluster $m$ includes  $L$ APs, $K$ UEs and a dedicated local processing unit. The APs are equipped with $N$ antennas, while the UEs have a single antenna.. In this paper, we identify the local processing units within each cluster in order to devise the precoder for UEs by considering the local CSI obtained from all APs in the cluster. In relation to the model elucidated in this paper, it is imperative to acknowledge that a cluster is delineated as a distinct spatial domain characterized by pre-established dimensions that encapsulate a part of the coverage area. This cluster encompasses a predetermined quantity of APs and UEs alongside a local processing unit devoted to the formulation of the precoders. Through the systematic accumulation of a specific count of these clusters, the comprehensive coverage area can be diligently constructed.

\begin{figure}[h]
\raggedright
\includegraphics[width=9cm]{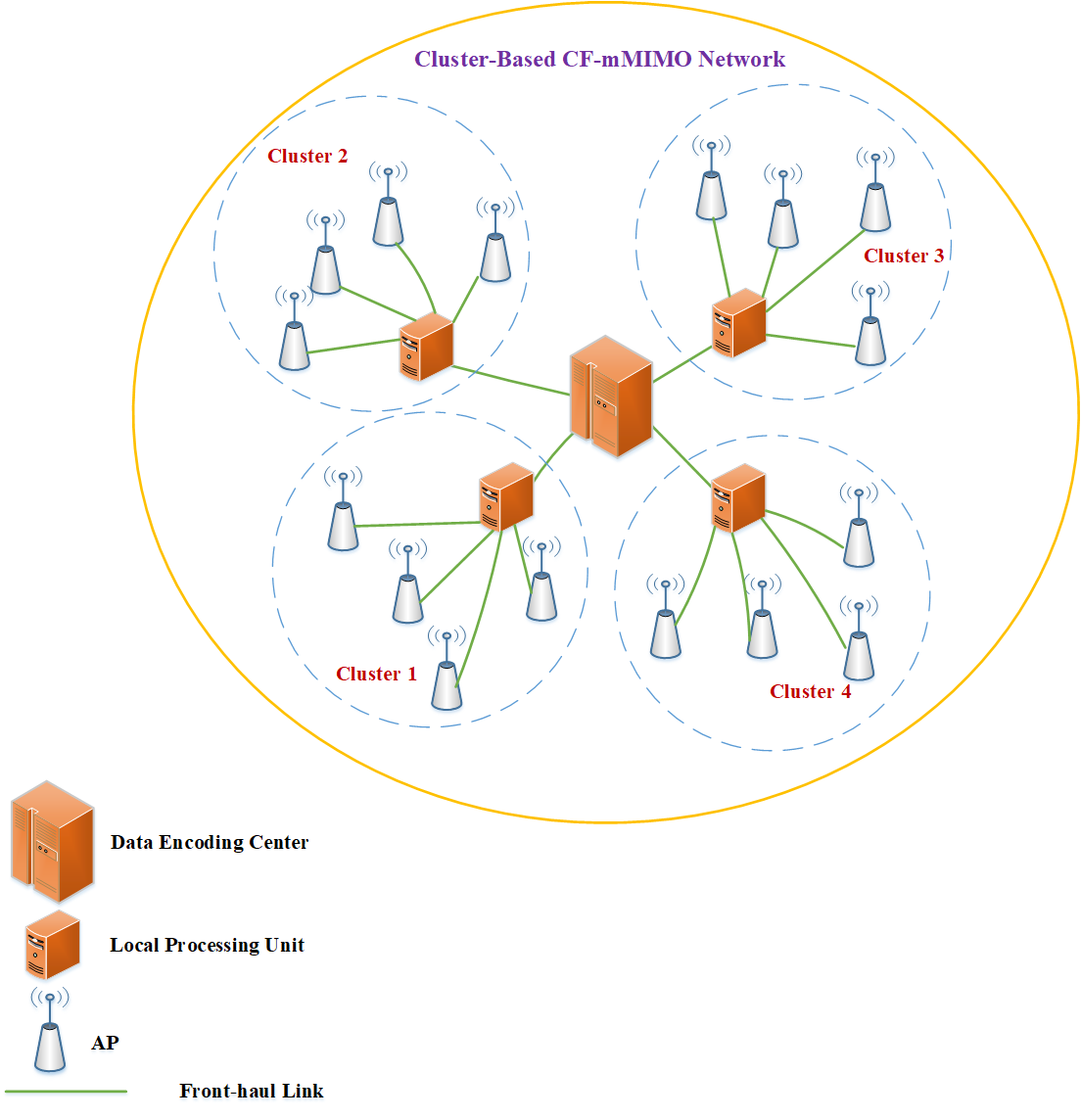}
\caption{The main components of a cluster-based CF-mMIMO system.}
\label{figure one}
\end{figure}

In our model, it is presumed that all cluster APs are interconnected with the local processing unit through front-haul links, and perfect CSI is accessible in all the local processing units. Furthermore, we take into account the presence of a front-haul link connecting the local processing unit and the data encoding center, facilitating the transmission of encoded signals to all the clusters.
Figure \ref{figure one} illustrates the primary constituents of a cluster-based CF-mMIMO system. It is important to note that, in this paper, we adopt the assumption that all clusters cater to all UEs present within the network.

\subsection{Cluster-Base CF-mMIMO DL Operation}

In this paper, we assume each cluster serves all $K_T$ UEs in the same time-frequency resources. The received downlink signal at UE $k$ is

\begin{align}\label{DL received signal}
y_k^{dl} =& \sum\limits_{m = 1}^M {\sum\limits_{l = 1}^L {{\bf{h}}_{kl,m}^{\rm{H}}} } {{\bf{x}}_{l,m}} + {n_k}\nonumber\\ =& \sum\limits_{m = 1}^M {\sum\limits_{l = 1}^L {{\bf{h}}_{kl,m}^{\rm{H}}} } \left( {\sum\limits_{i = 1}^{{K_T}} {{{\bf{w}}_{il,m}}{s_i}} } \right) + {n_k}
\end{align}
where ${n_{k}} \sim \mathcal{N_{C}}(0,{\sigma^2})$ is the receiver noise and

\begin{equation}\label{transmitted DL signal}
{{\bf{x}}_{l,m}} = \sum\limits_{i = 1}^{{K_T}} {{{\bf{w}}_{il,m}}{s_i}} 
\end{equation}
is the transmitted signal by AP $l$ in cluster $m$. This signal is generated as the sum of the 
UEs' signals where each term intended for UE $i$ contain the zero mean and unit-power downlink data signal $s_i \in \rm{C}$ ( with ${\rm{E\{ }}{\left| {{s_i}} \right|^2}\}  = 1$) and the transmit precoder ${{\bf{w}}_{il,m}}$.
Utilizing (\ref{transmitted DL signal}), we rewrite (\ref{DL received signal}) in compact form as:

\begin{align}\label{compact DL received signal}
y_k^{dl} = &{\sum\limits_{m = 1}^M {\sum\limits_{i = 1}^{{K_T}} {\left[ {\begin{array}{*{20}{c}}
{{{\bf{h}}_{k1,m}}}\\
 \vdots \\
{{{\bf{h}}_{kL,m}}}
\end{array}} \right]} } ^{\rm{H}}}\left[ {\begin{array}{*{20}{c}}
{{{\bf{w}}_{i1,m}}}\\
 \vdots \\
{{{\bf{w}}_{iL,m}}}
\end{array}} \right]{s_i} + {n_k} \nonumber\\= &\sum\limits_{m = 1}^M {\sum\limits_{i = 1}^{{K_T}} {{\bf{h}}_{k,m}^{\rm{H}}{{\bf{w}}_{i,m}}{s_i} + } } {n_k}
\end{align}
where ${{\bf{h}}_{k,m}} = {\left[ {{\bf{h}}_{k1,m}^{\rm{T}}, \cdots ,{\bf{h}}_{kL,m}^{\rm{T}}} \right]^{\rm{T}}}\in {{\rm{C}}^{LN \times 1}}$  denotes the collective channel to UE $k$ from all APs in the cluster $m$, and ${{\bf{w}}_{i,m}} = {\left[ {{\bf{w}}_{i1,m}^{\rm{T}}, \cdots ,{\bf{w}}_{iL,m}^{\rm{T}}} \right]^{\rm{T}}} \in {{\rm{C}}^{LN \times 1}}$ denotes the collective precoder assigned to UE $i$ by local processing unit in cluster $m$.

The system model formulation in (\ref{compact DL received signal}) provides a comprehensive perspective on the DL transmission at the network level, and is particularly useful when characterizing the cluster-based operation. We assume, in each arbitrary cluster $m \in \{ 1, \cdots ,M\} $, the local processing unit  accesses the collective channels $\left\{ {\left. {{{\bf{h}}_{k,m}}:k = 1, \cdots ,{K_T}} \right\}} \right.$ in the DL. These collective channels are applied by the local processing unit to compute the collective precoders $\left\{ {\left. {{{\bf{w}}_{k,m}}:k = 1, \cdots ,{K_T}} \right\}} \right.$ for all $K_T$ UEs in the network. On the other hand, the DL data $\left\{ {\left. {{s_i}:i = 1, \cdots ,{K_T}} \right\}} \right.$ encoded by the data encoding center, transmit to all local processing units in the network, and the local processing unit $m \in \{ 1, \cdots ,M\} $, generate the signal ${{\bf{x}}_{l,m}}$ in (\ref{transmitted DL signal}) for each AP using the percoder and the DL data. So we can express that in a cluster-based CF-mMIMO system,  the DL signal processing tasks, including data encoding and precoding, are divided between the data encoding center in the network and the local processing units in each cluster. APs perform no signal processing functions in this system. The specific role of APs is to relay the precoded data symbols to all $K_T$ UEs in the network.

\begin{figure}[h]
\raggedright
\includegraphics[width=9cm]{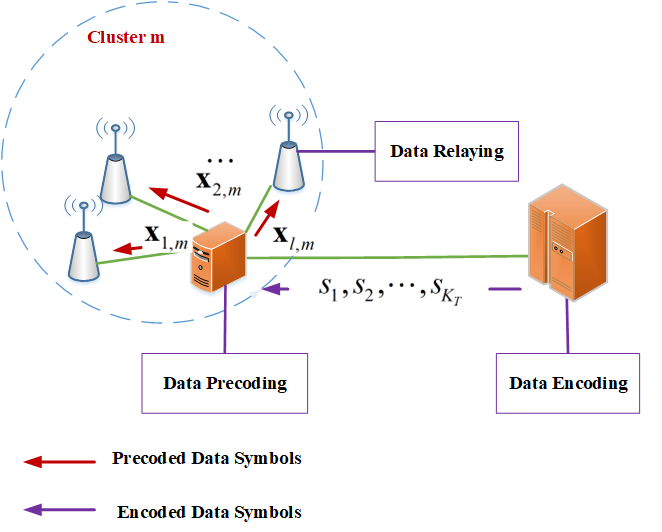}
\caption{The DL signal processing operation in a cluster-based CF-mMIMO.}
\label{figure two}
\end{figure}
Figure \ref{figure two} illustrates the DL signal processing operation of a cluster-based CF-mMIMO system.

\subsection{Spectral Efficiency in cluster-based CF-mMIMO Systems}

We will now provide an achievable SE expression that applies when using any type of precoder.

{\bf{Corollary}}: The achievable SE of UE $k$ in the DL in a cluster-based CF-mMIMO system is lower bounded by:

\begin{equation}
{\rm{S}}{{\rm{E}}_k} = {\log _2}(1 + {\rm{SIN}}{{\rm{R}}_k})
\end{equation}
with the effective SINR given by

\begin{align}
&{\rm{SIN}}{{\rm{R}}_k} =\nonumber\\
&\frac{{{{\left| {\sum\limits_{m = 1}^M {{\rm{E}}\left\{ {\left. {{\bf{h}}_{k,m}^{\rm{H}}{{\bf{w}}_{k,m}}} \right\}} \right.} } \right|}^2}}}{{\sum\limits_{i = 1}^K {{\rm{E}}\left\{ {\left. {{{\left| {\sum\limits_{m = 1}^M {{\bf{h}}_{k,m}^{\rm{H}}{{\bf{w}}_{i,m}}} } \right|}^2}} \right\} - {{\left| {\sum\limits_{m = 1}^M {{\rm{E}}\left\{ {\left. {{\bf{h}}_{k,m}^{\rm{H}}{{\bf{w}}_{k,m}}} \right\}} \right.} } \right|}^2} + {\sigma ^2}} \right.} }}
\end{align}

{\bf{Proof}}: Since in each cluster, only the knowledge of the average of the effective channels ${{\rm{E}}\left\{ {\left. {{\bf{h}}_{k,m}^{\rm{H}}{{\bf{w}}_{k,m}}} \right\}} \right.}$ is known by the UE $k$, the received signal in (\ref{compact DL received signal}) can be expressed as

\begin{align}\label{DL signal for corollary proof}
y_k^{dl} =& \left( {\sum\limits_{m = 1}^M {{\rm{E}}\left\{ {\left. {{\bf{h}}_{k,m}^{\rm{H}}{{\bf{w}}_{k,m}}} \right\}} \right.} } \right){s_k}\nonumber\\
& + \left( {\sum\limits_{m = 1}^M {\left( {{\bf{h}}_{k,m}^{\rm{H}}{{\bf{w}}_{k,m}} - {\rm{E}}\left\{ {\left. {{\bf{h}}_{k,m}^{\rm{H}}{{\bf{w}}_{k,m}}} \right\}} \right.} \right)} } \right){s_k}\\
& + \sum\limits_{i = 1,i \ne k}^K {\sum\limits_{m = 1}^M {{\bf{h}}_{k,m}^{\rm{H}}{{\bf{w}}_{i,m}}{s_i} + } } {n_k}\nonumber
\end{align}
which is a deterministic channel with additive interference term

\begin{align}
i_{k} =& \left( {\sum\limits_{m = 1}^M {\left( {{\bf{h}}_{k,m}^{\rm{H}}{{\bf{w}}_{k,m}} - {\rm{E}}\left\{ {\left. {{\bf{h}}_{k,m}^{\rm{H}}{{\bf{w}}_{k,m}}} \right\}} \right.} \right)} } \right){s_k}\nonumber\\ &+ \sum\limits_{i = 1,i \ne k}^K {\sum\limits_{m = 1}^M {{\bf{h}}_{k,m}^{\rm{H}}{{\bf{w}}_{i,m}}{s_i}} } 
\end{align}
and additive noise $n_k$. The $i_k$ has zero mean (since $s_k$ is zero mean) and although it consists of the desired signal $s_k$, it is uncorrelated with it since

\begin{equation}
\begin{array}{l}
{\rm{E}}\left\{ {\left. {s_k^*{i_k}} \right\} = } \right.\\
\underbrace {{\rm{E}}\left\{ {\left. {\left( {\sum\limits_{m = 1}^M {\left( {{\bf{h}}_{k,m}^{\rm{H}}{{\bf{w}}_{k,m}} - {\rm{E}}\left\{ {\left. {{\bf{h}}_{k,m}^{\rm{H}}{{\bf{w}}_{k,m}}} \right\}} \right.} \right)} } \right)} \right\}} \right.}_{ = 0}{\rm{E}}\left\{ {\left. {{{\left| {{s_k}} \right|}^2}} \right\} = 0} \right.
\end{array}
\end{equation}
By noting that the signals of different UEs are independent, we have

\begin{align}
{\rm{E}}\left\{ {\left. {{{\left| {{i_k}} \right|}^2}} \right\}} \right.& = \sum\limits_{i = 1}^K {{\rm{E}}\left\{ {\left. {{{\left| {\sum\limits_{m = 1}^M {{\bf{h}}_{k,m}^{\rm{H}}{{\bf{w}}_{i,m}}} } \right|}^2}} \right\}} \right.}\nonumber\\& - {{\left| {\sum\limits_{m = 1}^M {{\rm{E}}\left\{ {\left. {{\bf{h}}_{k,m}^{\rm{H}}{{\bf{w}}_{k,m}}} \right\}} \right.} } \right|}^2} 
\end{align}
following Lemma 3.3 on p.244 in \cite{demir2021foundations} the proof completes.

\section{cluster-based Transmit Precoding}\label{section three}

In this section, we discuss the structure of the DL transmit precoders for a cluster-based CF-mMIMO system. In a general form, we define ${{\bf{w}}_{k,m}}$ as follows:

\begin{equation}\label{general precoder form}
{{\bf{w}}_{k,m}} = \sqrt {{\rho _{k,m}}} \frac{{{{{\bf{\bar w}}}_{k,m}}}}{{\sqrt {{\rm{E}}\left\{ {\left. {{{\left\| {{{{\bf{\bar w}}}_{k,m}}} \right\|}^2}} \right\}} \right.} }}
\end{equation}
where ${\rho _{k,m}}$ is the total transmit power assigned to UE $k$ from all APs in cluster $m$ and ${{{\bf{\bar w}}}_{k,m}} \in {{\rm{C}}^{LN \times 1}}$ is the normalized precoder vector that points out the direction of precoder. Now,  we propose two forms of precoders for cluster-based CF-mMIMO systems in the following two parts.

\begin{enumerate}
\item MMSE Precoding: The MMSE precoder is obtained from (\ref{general precoder form}) using

\begin{align}\label{MMSE precoder}
{\bf{\bar w}}_{k,m}^{{\rm{MMSE}}} = {p_k}{\left( {\sum\limits_{i = 1}^{{K_T}} {{p_i}{{\bf{h}}_{i,m}}{\bf{h}}_{i,m}^{\rm{H}}}  + {\sigma ^2}{{\bf{I}}_{NL}}} \right)^{ - 1}}{{\bf{h}}_{k,m}}
\end{align}
where $p_i$ is the uplink power of UE $i$ and ${\bf I}$ is the identity matrix. Intuitively, by using MMSE precoders, we can transmit a strong signal to the desired UE and limit the interference caused to other UEs. 

\item MR precoding: This precoder is achieved form (\ref{general precoder form})  using

\begin{equation}
{\bf{\bar w}}_{k,m}^{{\rm{MR}}} = {{\bf{h}}_{k,m}}
\end{equation}
This scheme maximizes the numerator of the effective SINR (i.e. the fraction of the transmit power from cluster $m$ that is received at the desired UE). However, MR precoder ignores the interference that the cluster is causing among the all UEs in the network.
\end{enumerate}
In the subsequent section, we analyze and contrast the computational complexity of MMSE precoders in three distinct systems: cluster-based, centralized , and distributed CF-mMIMO systems.

\section{Computational Complexity Analysis}\label{section four}
To facilitate a comparative analysis of the computational complexity of MMSE precoders in cluster-based architecture against both centralized and distributed architectures, it is imperative to consider the relationship between the channel of UE $k$ in the centralized architecture ${\bf h}_k$ and the channel between  UE $k$ and AP $q$ in the distributed architecture ${\bf h}_{kq}$ with the channels associated with UEs in the cluster-based architecture. The relationship between the channels is as follows:

\begin{equation}
{{\bf{h}}_k} = {\left[ {\begin{array}{*{20}{c}}
{{\bf{h}}_{k,1}^{\rm{T}}}& \cdots &{{\bf{h}}_{k,M}^{\rm{T}}}
\end{array}} \right]^{\rm{T}}} \in {{\rm{C}}^{NLM \times 1}}
\end{equation}
\begin{equation}
{{\bf{h}}_{kq}} = {{\bf{h}}_{kl,m}} \in {{\rm{C}}^{N \times 1}},\,\,q \in \left\{ {\begin{array}{*{20}{c}}
{1,}&{ \cdots ,}&{ML}
\end{array}} \right\}\,\,\,
\end{equation}
Following \cite{demir2021foundations} the normalized MMSE precoders for centralized and distributed architecture express as follows:
\begin{equation}
{\bf{\bar w}}_k^{{\rm{MMSE,Cent}}{\rm{.}}} = {p_k}{\left( {\sum\limits_{i = 1}^{{K_T}} {{p_i}{{\bf{h}}_i}{\bf{h}}_i^{\rm{H}}}  + {\sigma ^2}{{\bf{I}}_{NLM}}} \right)^{ - 1}}{{\bf{h}}_k}
\end{equation}
\begin{equation}
{\bf{\bar w}}_{kq}^{{\rm{MMSE,Dist}}.} = {p_k}{\left( {\sum\limits_{i = 1}^{{K_T}} {{p_i}{{\bf{h}}_{iq}}} {\bf{h}}_{iq}^{\rm{H}} + {\sigma ^2}{{\bf{I}}_N}} \right)^{ - 1}}{{\bf{h}}_{kq}}
\end{equation}
The computational complexity of determining the MMSE precoders for centralized, distributed, and cluster-based cell-free massive MIMO systems can be evaluated utilizing the methodology outlined in Appendix B.1.1 of \cite{bjornson2017massive}, . In Table \ref{table 1}, a comprehensive overview is presented, illustrating the total quantity of complex multiplications necessary to compute the MMSE precoder for UE $k$ under three distinct cell-free massive MIMO system architectures.

\begin{table} [h!]
\caption{Computational Complexity of MMSE Precoders in Cluster-Based CF-mMIMO Compared with Centralized and Distributed CF-mMIMO}
\label{table 1}
\begin{center}
\begin{tabular}{||c|c||} 
 \hline
  Scheme & MMSE Precoder Vector Computation \\ [0.1ex] 
 \hline\hline
 
  Centralized & $\frac{{{{(NLM)}^2} + NLM}}{2}{K_T} + {(NLM)^2} + \frac{{{{(NLM)}^3} - NLM}}{3}$ \\ 
 \hline
  Distributed & $\frac{{{N^2} + N}}{2}{K_T}{L_T} + {N^2}{L_T} + \frac{{{N^3} - N}}{3}{L_T}$  \\
 \hline
   Cluster-Based &  $\frac{{{{(NL)}^2} + NL}}{2}{K_T}M + {(NL)^2}M + \frac{{{{(NL)}^3} - NL}}{3}M$ \\
 \hline
\end{tabular}
\end{center}
\end{table}

Based on the preceding discussion, it is evident that the centralized architecture is characterized by high complexity due to the necessity of inverting an $NLM \times NLM$ matrix. In contrast, the cluster-based architecture falls between the complexities of the centralized and distributed architectures.

\section{Numerical Results}\label{section five}
This section presents the numerical results comparing the performance of the proposed cluster-based architecture of a CF-mMIMO system with its centralized and distributed architectures in terms of computational complexity and the SE of UEs in the network. In our simulation, we employ a square region measuring $980m \times 980m$ and partition this area into $ M = 2, 4, 8$ clusters, each of equal size. Subsequently, we randomly distribute $ L_T = 96$ APs and $ K_T =40$ UEs within this spatial domain, adhering to a uniform distribution pattern. Consequently, it is plausible to assert that within the target network, the APs are deployed at an average separation distance of $100m$ from one another. We have assumed that all clusters have an equal number of access points and users, which is consistent with the uniform distribution of UEs and APs.

\begin{figure}[h]
\raggedleft
\includegraphics[width=6cm]{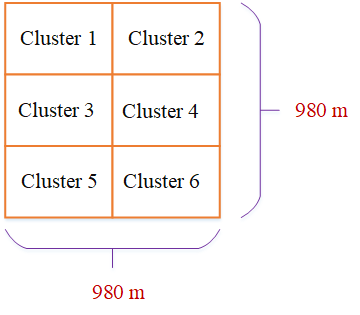}
\caption{The schematic of simulation area for $M=3 \times 2$ clusters.}
\label{figure three}
\end{figure}

Figure (\ref{figure three}) depicts a schematic representation of the simulation area corresponding to $M = 3 \times 2$ clusters. In order to evaluate the performance of the cluster-based structure, we conduct a comparative analysis involving two alternative architectures: centralized and distributed structures. These architectures are compared to the cluster-based network that has been previously presented. In both the centralized and distributed frameworks, we take into account all APs and UEs within the cluster-based network. However, there is a distinction between the two structures: the centralized structure assumes that the precoder for each user is determined at the CPU based on the global CSI obtained from the entire network, while the distributed structure acquires the precoder based on the local CSI at each AP.

\begin{table} [h!]
\caption{Simulation Parameters for Cluster-Based CF-mMIMO System}
\label{table 2}
\begin{center}

\begin{tabular}{|c@{\hspace{6mm}}|c@{\hspace{6mm}}|}
    \hline
    	 Area & $980m \times 980m$  \\
    	 Bandwidth & $20 MHz$ \\
         Total number of APs in the network & $L_T = 96$ \\
         Total number of UEs in the network & $K_T = 40$\\
         Number of Antenna per APs & $N = 4$  \\
         Pathloss Exponent  & $\alpha = 3.76$ \\
         Maximum DL Transmit Power per AP & $1000 mW$ \\
         Noise Power & $-94 dBm$ \\
         Standard Deviation of Shadowing & $\sigma_{sf} = 4 dB$ \\
         UP-Link Transmit Power & $p_k = 100 mW$ \\
    \hline      
\end{tabular}

\end{center}
\end{table}

The large-scale fading coefficients and channel realizations are generated as \cite{bjornson2019making}

\begin{equation}
{\beta _{kl,m}} =  - 30.5 - 37.6{\log _{10}}(\frac{{{d_{kl,m}}}}{{1m}}) + {\alpha _{kl,m}}\,\,\,\,\,{\rm{dB}}
\end{equation}

\begin{equation}
{{\bf{h}}_{kl,m}} \sim \mathcal{N_{C}}(0, {\beta _{kl,m}}{\bf{I}}_N)
\end{equation}
where ${d _{kl,m}}$ is the distance of AP $l$ in cluster $m$ from UE $k$ in the network. 
The APs are positioned $10m$ above the UEs, and these distances are taken into account when calculating the distances. The shadow fading is modeled by a Gaussian distribution with a mean of zero and a standard deviation of $\sigma_{sf} = 4 dB$, represented as ${\alpha_{k}} \sim \mathcal{N_{C}}(0,{4^2})$. All other simulation parameters are documented in table \ref{table 2}. Monte Carlo simulations were conducted employing 100 randomly generated sets of AP and UE locations for each individual figure. Following the aggregation of results from 300 distinct channel realizations for each set up, the graphical representations were derived by means of evaluating the SE expression as stipulated in the Corollary.

\subsection{Complexity Comparison}
In this section, we perform a numerical analysis to assess the computational complexity entailed in designing the MMSE precoder within both cluster-based and centralized  configurations.  

\begin{figure}[h]
\raggedright
\includegraphics[width=9cm]{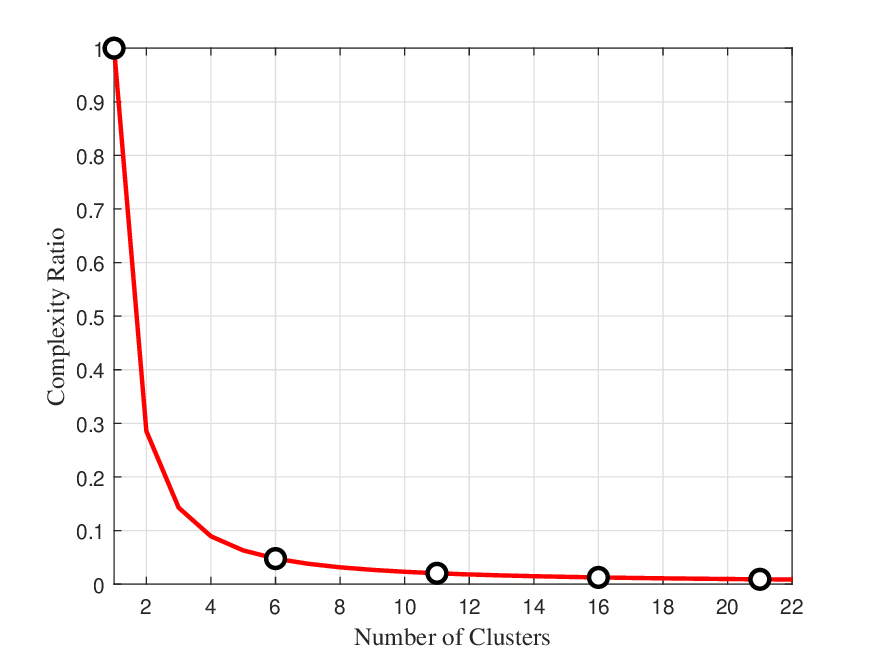}
\caption{The relative complexity between cluster-based and centralized structures in relation to the number of clusters.}
\label{figure six}
\end{figure}

Figure (\ref{figure six}) quantifies the computational complexity discrepancy between the cluster-based structure and the centralized structure. The horizontal axis of the figure represents the number of clusters utilized in network construction, while the vertical axis depicts the ratio of the computational complexity associated with the cluster-based structure to that of the centralized structure when computing the MMSE percoder.

As illustrated the Figure (\ref{figure six}), it is apparent that constructing a CF-mMIMO network using a cluster-based structure leads to a reduction in the computational complexity necessary for MMSE precoder design, as the number of clusters increases, as compared to the corresponding centralized structure.

\begin{table} [h!]
\caption{The computational complexity associated with the cluster-based structure, given a specific number of clusters, and the corresponding centralized structure in MMSE precoder design.}
\label{table 3}
\begin{center}
\begin{tabular}{|l|c|c|c|c|c|}
  \hline
  \diagbox[width=8.5em,trim=r]{Scheme}{Cluster Num.} & 1 & 2 & 4 & 8 & 16 \\
  \hline
  Centralized & $2.1e^7$ & $2.1e^7$ & $2.1e^7$ & $2.1e^7$ & $2.1e^7$ \\
  \hline
  Cluster-Based & $2.1e^7$ & $6.2e^6$ & $1.9e^6$ & $6.8e^5$ & $2.7e^5$ \\
   \hline
  Complexity Ratio & 1 & 0.285 & 0.089 & 0.031 & 0.012 \\
  \hline
 \end{tabular}
\end{center}
\end{table}

Table \ref{table 3} presents the quantitative measurement of the computational complexity involved in the MMSE precoder design within a cluster-based structure. Specifically, the table showcases the values corresponding to different cluster configurations, namely $M = 1, 2, 4, 8, 16$ clusters. Furthermore, the table also encompasses the computational complexity associated with the corresponding centralized structure for comprehensive comparison.

Based on the findings derived from Figure (\ref{figure six}) and Table \ref{table 3}, there is evident support indicating that an increased number of clusters in the design of a cluster-based CF-mMIMO network results in reduced computational complexity.

\subsection{SE Comparison}
In this section, we will conduct a numerical analysis to compare the SE of the cluster-based architectural framework with two distinct centralized and distributed structural configurations. This evaluation will be carried out under the condition that the cluster-based system is configured with $ M = 2 \times 1$, $ M = 2 \times 2$ and $ M = 2 \times 4$ clusters.

\begin{figure}[h]
\raggedright
\includegraphics[width=9cm]{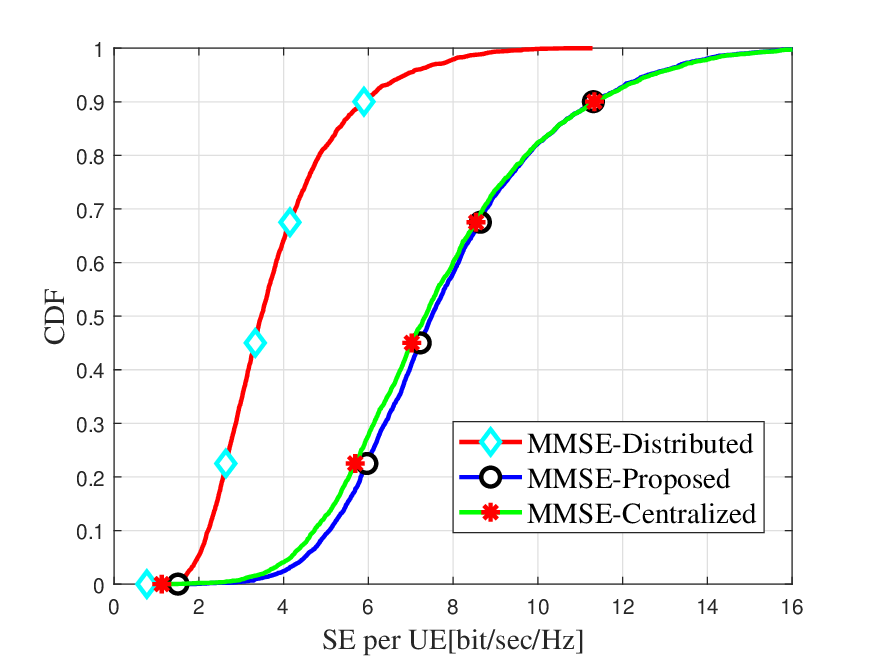}
\caption{CDF of DL SE per UE with MMSE precoder for $M = 2 \times 1$ clusters in a cluster-based CF-mMIMO system and comparison with corresponding centralized and distributed architectures.}
\label{figure five}
\end{figure}

\begin{figure}[h]
\raggedright
\includegraphics[width=9cm]{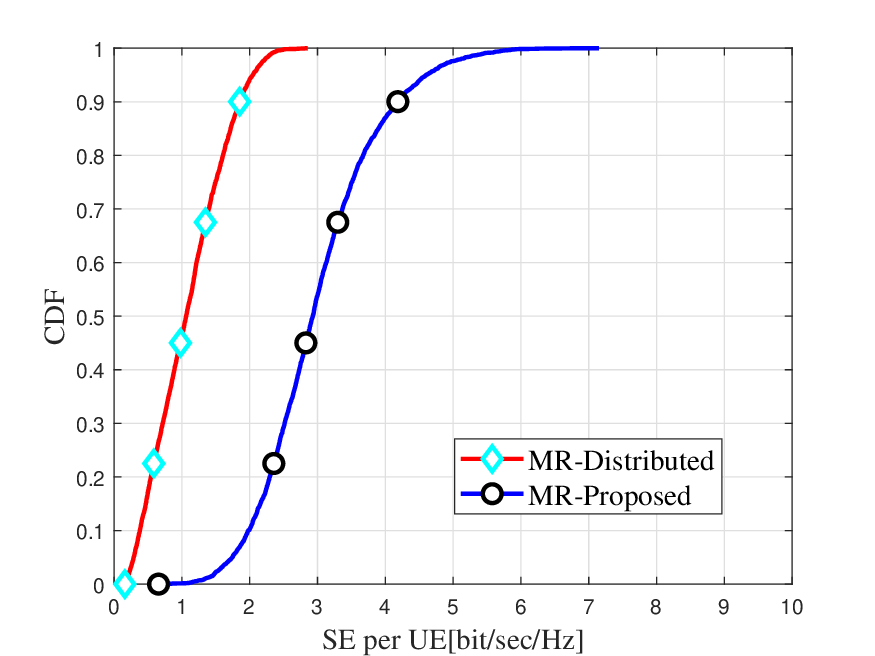}
\caption{CDF of DL SE per UE with MR precoder for $M = 2 \times 1$ clusters in a cluster-based CF-mMIMO system and comparison with corresponding centralized and distributed architectures. }
\label{figure eight}
\end{figure}
Figures \ref{figure five} and \ref{figure eight} present a comparative analysis of the spectral efficiency between the cluster-based architectural configuration and its two centralized and distributed structural counterparts. In both figures, the cluster-based structure is assumed to be configured with 2 clusters. As distinctly evident from the data presented in these two figures, configuring the cluster-based structure with two clusters results in a notable enhancement of users SE when compared to the distributed structure employing both MR and MMSE precoders. The principal enhancement of the cluster-based architecture, in relation to its capacity to elevate SE, is prominently demonstrated through the implementation of the system with the  MMSE precoders. In this context, as illustrated in Figure \ref{figure five}, the cluster-based configuration is capable of achieving relatively superior performance compared to the centralized counterpart. Importantly, this performance gain is achieved concomitant with a substantial reduction in computational complexity, as elucidated in the preceding section.
This phenomenon can be justified by the fact that when a cluster-based system is designed using MMSE precoder with 2 clusters, the number of APs within each cluster equals half the total number of APs in the network. Consequently, this arrangement has the effect of augmenting the power of the MMSE precoder in each cluster, resulting in increased power and diminished interference. While the precoder designed for one cluster does not influence the interference caused by another cluster, it is important to note that the magnitude of this interference remains negligible due to the large number of APs within each cluster. Moreover, considering the network's limited two-cluster configuration, the number of interfering factors remains relatively low. As a result, the overall level of interference experienced by each user ultimately proves to be minimal.

\begin{figure}[h]
\raggedright
\includegraphics[width=9cm]{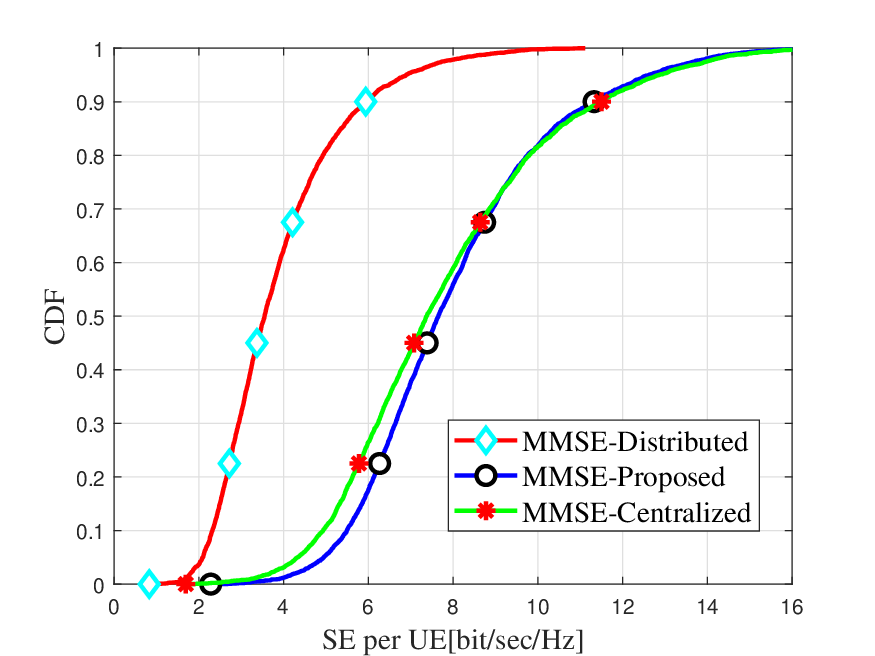}
\caption{CDF of DL SE per UE with MMSE precoder for $M = 2 \times 2$ clusters in a cluster-based CF-mMIMO system and comparison with corresponding centralized and distributed architectures. }
\label{figure four}
\end{figure}

\begin{figure}[h]
\raggedright
\includegraphics[width=9cm]{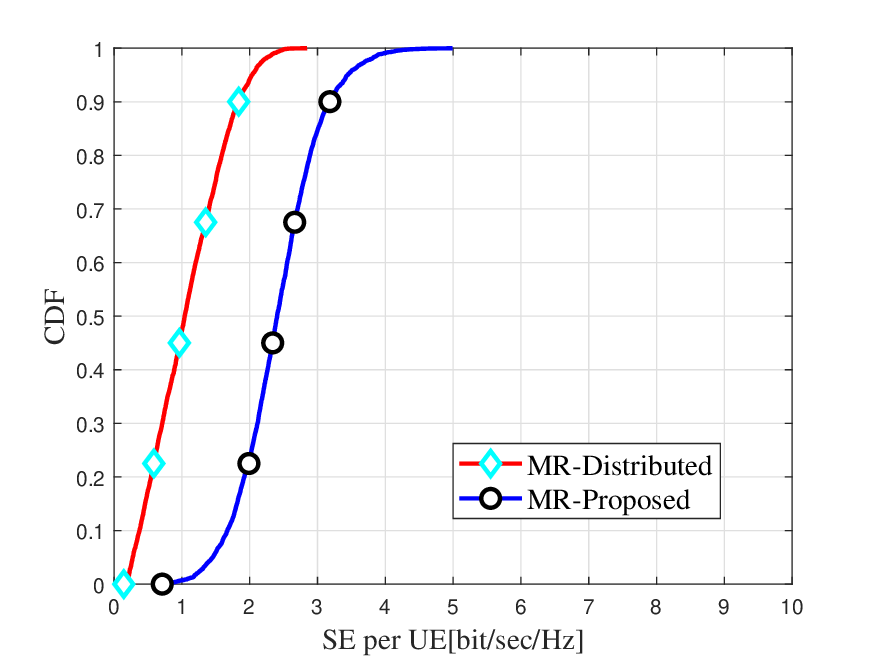}
\caption{CDF of DL SE per UE with MR precoder for $M = 2 \times 2$ clusters in a cluster-based CF-mMIMO system and comparison with corresponding centralized and distributed architectures. }
\label{figure nine}
\end{figure}

Figures \ref{figure four} and \ref{figure nine} provide a performance evaluation of the cluster-based configuration featuring 4 clusters in comparison to two distinct centralized and distributed structural setups with respect to spectral efficiency. A comparative analysis between Figure \ref{figure nine} and Figure \ref {figure eight} allows us to draw the inference that, despite the decrease in performance of the cluster-based structure employing the MR precoder as the number of clusters increases from 2 to 4, the overall performance in this context still surpasses that of the distributed structure. The decline in performance of the cluster-based configuration using the MR precoder under these conditions can be attributed to the escalation in the number of interfering factors. As the number of clusters increases, the interference per user intensifies. With regard to the performance of the cluster-based configuration featuring 4 clusters and employing the MMSE precoder, a comparative analysis between Figures \ref{figure four} and \ref{figure five} reveals that as the number of clusters increases from 2 to 4, there is an observable enhancement in system performance. This improvement can be attributed to the augmented efficacy of the MMSE precoder in augmenting signal power and mitigating interference within the network.

\begin{figure}[h]
\raggedright
\includegraphics[width=9cm]{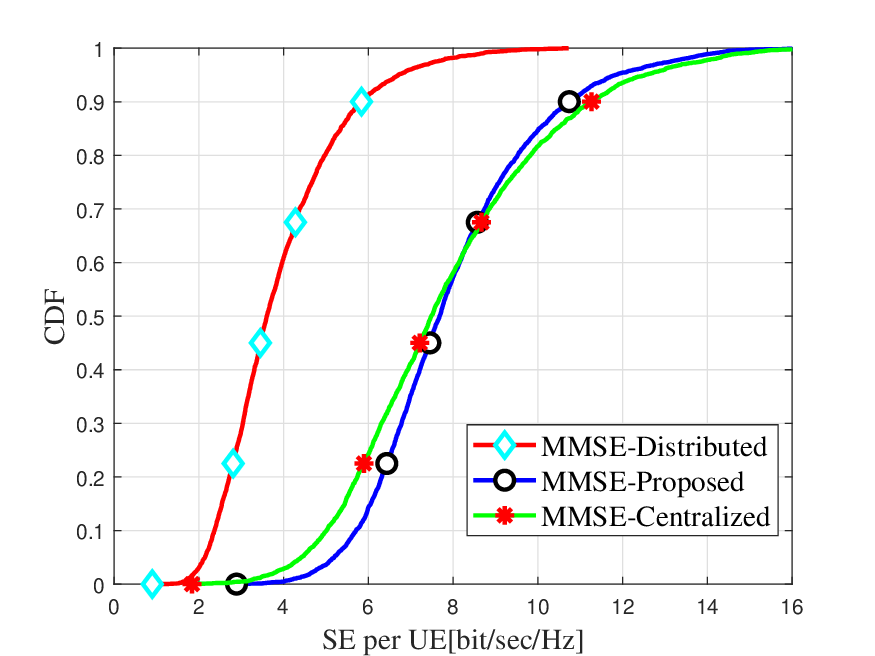}
\caption{CDF of DL SE per UE with MMSE precoder for $M = 2 \times 4$ clusters in a cluster-based CF-mMIMO system and comparison with corresponding centralized and distributed architectures. }
\label{figure seven}
\end{figure}

\begin{figure}[h]
\raggedright
\includegraphics[width=9cm]{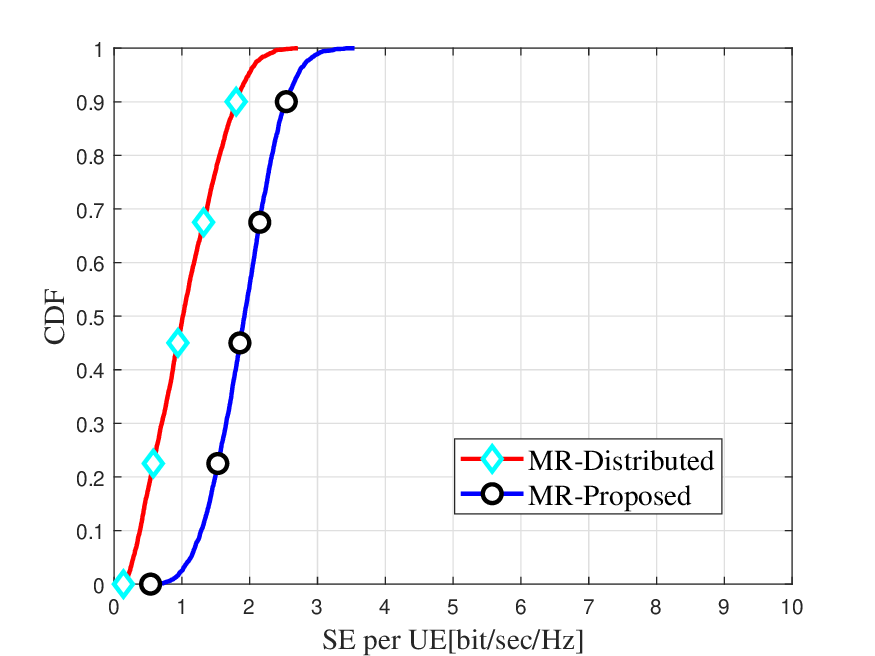}
\caption{CDF of DL SE per UE with MR precoder for $M = 2 \times 4$ clusters in a cluster-based CF-mMIMO system and comparison with corresponding centralized and distributed architectures. }
\label{figure ten}
\end{figure}
In our final analysis, as depicted in Figures \ref{figure seven} and \ref{figure ten}, we scrutinize the performance of the cluster-based architecture when configured with 8 clusters. In a broader context, it becomes evident that the cluster-based structure employing 8 clusters outperforms the distributed structure with both MR and MMSE precoders. However, it is noteworthy that the performance of this configuration, when contrasted with the design mode featuring 4 clusters, experiences a marginal decline. In summary of the aforementioned findings, it can be inferred that within the cluster-based architectural framework, the system's performance, particularly in the context of spectral efficiency when utilizing the MMSE precoder, attains its zenith when the structure is configured with 4 clusters. Under this configuration, it becomes feasible to achieve performance levels that are comparable or slightly superior to those of the centralized structure, all while benefiting from a significantly reduced computational complexity.

\subsection{Subsection Heading Here}
Subsection text here.


\subsubsection{Subsubsection Heading Here}
Subsubsection text here.

\section{Conclusion}\label{section six}
In this paper, a novel architectural framework termed the "cluster-based structure" was introduced for CF-mMIMO systems, leveraging a combination of the preexisting centralized and distributed structures. This innovative structure, facilitated by the implementation of a unique cluster configuration and the design of the MMSE precoder tailored to this cluster model, demonstrates the capacity to substantially enhance the network's SE in comparison to the distributed structure. Simultaneously, it achieves a significant reduction in computational complexity when compared to the centralized structure. The simulation results demonstrate that the cluster-based structure outperforms the distributed structure in terms of enhancing the SE of network UEs, with both MR and MMSE precoders. Furthermore, the behavior of the cluster-based structure can be altered by adjusting the number of clusters, resulting in changes in the SE of the network. Specifically, if the system is implemented with an MR precoder, the spectral efficiency decreases as the number of clusters increases. However, if the network is designed with four clusters and an MMSE precoder is used, the computational complexity can be reduced by $85\%$ compared to the centralized structure while achieving maximum SE.

\section{Appendix}
According to (\ref{MMSE precoder}) we define {\bf A} and {\bf B} as follows:

\begin{equation}
{\bf{A}} = \sum\limits_{i = 1}^{{K_T}} {{p_i}{{\bf{h}}_{i,m}}{\bf{h}}_{i,m}^{\rm{H}}}  + {\sigma ^2}{{\bf{I}}_{LN}} \in {{\rm{C}}^{NL \times NL}}
\end{equation}

\begin{equation}
{\bf{B}} = {p_k}{{\bf{h}}_{k,m}} \in {{\rm{C}}^{NL \times 1}}
\end{equation}
By utilizing Lemma B.1 and B.2 of \cite{bjornson2017massive}, the computational requirements within each cluster can be determined. Specifically, to compute ${\bf A^{-1}}$, ${\bf A^{-1}}{\bf B}$ and ${\sum\limits_{i = 1}^{{K_T}} {{p_i}{{\bf{h}}_i}{\bf{h}}_i^{\rm{H}}} }$, the following number of complex multiplications are needed: $\frac{{{{(NL)}^3} - NL}}{3}$, $(NL)^2$  and  $\frac{{{{(NL)}^2} + NL}}{2}{K_T}$, respectively. As a result, the total number of complex multiplications across all $M$ clusters can be obtained as follows:

\begin{equation}
\frac{{{{(NL)}^2} + NL}}{2}{K_T}M + {(NL)^2}M + \frac{{{{(NL)}^3} - NL}}{3}M
\end{equation}

\appendices

\ifCLASSOPTIONcaptionsoff
  \newpage
\fi

\bibliographystyle{ieeetr}
\bibliography{bibRef}

\begin{thebibliography}{10}

\bibitem{ngo2017cell}
H.~Q. Ngo, A.~Ashikhmin, H.~Yang, E.~G. Larsson, and T.~L. Marzetta,
  ``Cell-free massive mimo versus small cells,'' {\em IEEE Transactions on
  Wireless Communications}, vol.~16, no.~3, pp.~1834--1850, 2017.

\bibitem{bjornson2019making}
E.~Bj{\"o}rnson and L.~Sanguinetti, ``Making cell-free massive mimo competitive
  with mmse processing and centralized implementation,'' {\em IEEE Transactions
  on Wireless Communications}, vol.~19, no.~1, pp.~77--90, 2019.

\bibitem{nayebi2017precoding}
E.~Nayebi, A.~Ashikhmin, T.~L. Marzetta, H.~Yang, and B.~D. Rao, ``Precoding
  and power optimization in cell-free massive mimo systems,'' {\em IEEE
  Transactions on Wireless Communications}, vol.~16, no.~7, pp.~4445--4459,
  2017.

\bibitem{zhang2019cell}
J.~Zhang, S.~Chen, Y.~Lin, J.~Zheng, B.~Ai, and L.~Hanzo, ``Cell-free massive
  mimo: A new next-generation paradigm,'' {\em IEEE Access}, vol.~7,
  pp.~99878--99888, 2019.

\bibitem{zhang2020prospective}
J.~Zhang, E.~Bj{\"o}rnson, M.~Matthaiou, D.~W.~K. Ng, H.~Yang, and D.~J. Love,
  ``Prospective multiple antenna technologies for beyond 5g,'' {\em IEEE
  Journal on Selected Areas in Communications}, vol.~38, no.~8, pp.~1637--1660,
  2020.

\bibitem{mai2018pilot}
T.~C. Mai, H.~Q. Ngo, M.~Egan, and T.~Q. Duong, ``Pilot power control for
  cell-free massive mimo,'' {\em IEEE Transactions on Vehicular Technology},
  vol.~67, no.~11, pp.~11264--11268, 2018.

\bibitem{akbar2018location}
N.~Akbar, S.~Yan, N.~Yang, and J.~Yuan, ``Location-aware pilot allocation in
  multicell multiuser massive mimo networks,'' {\em IEEE Transactions on
  Vehicular Technology}, vol.~67, no.~8, pp.~7774--7778, 2018.

\bibitem{buzzi2016survey}
S.~Buzzi, I.~Chih-Lin, T.~E. Klein, H.~V. Poor, C.~Yang, and A.~Zappone, ``A
  survey of energy-efficient techniques for 5g networks and challenges ahead,''
  {\em IEEE Journal on selected areas in communications}, vol.~34, no.~4,
  pp.~697--709, 2016.

\bibitem{demir2021foundations}
{\"O}.~T. Demir, E.~Bj{\"o}rnson, L.~Sanguinetti, {\em et~al.}, ``Foundations
  of user-centric cell-free massive mimo,'' {\em Foundations and
  Trends{\textregistered} in Signal Processing}, vol.~14, no.~3-4,
  pp.~162--472, 2021.

\bibitem{miretti2021team}
L.~Miretti, E.~Bj{\"o}rnson, and D.~Gesbert, ``Team precoding towards scalable
  cell-free massive mimo networks,'' in {\em 2021 55th Asilomar Conference on
  Signals, Systems, and Computers}, pp.~1222--1227, IEEE, 2021.

\bibitem{miretti2022team}
L.~Miretti, E.~Bj{\"o}rnson, and D.~Gesbert, ``Team mmse precoding with
  applications to cell-free massive mimo,'' {\em IEEE Transactions on Wireless
  Communications}, vol.~21, no.~8, pp.~6242--6255, 2022.

\bibitem{miretti2021precoding}
L.~Miretti, E.~Bj{\"o}rnson, and D.~Gesbert, ``Precoding for scalable cell-free
  massive mimo with radio stripes,'' in {\em 2021 IEEE 22nd International
  Workshop on Signal Processing Advances in Wireless Communications (SPAWC)},
  pp.~411--415, IEEE, 2021.

\bibitem{gouda2020distributed}
B.~Gouda, I.~Atzeni, and A.~T{\"o}lli, ``Distributed precoding design for
  cell-free massive mimo systems,'' in {\em 2020 IEEE 21st International
  Workshop on Signal Processing Advances in Wireless Communications (SPAWC)},
  pp.~1--5, IEEE, 2020.

\bibitem{atzeni2020distributed}
I.~Atzeni, B.~Gouda, and A.~T{\"o}lli, ``Distributed precoding design via
  over-the-air signaling for cell-free massive mimo,'' {\em IEEE Transactions
  on Wireless Communications}, vol.~20, no.~2, pp.~1201--1216, 2020.

\bibitem{bjornson2017massive}
E.~Bj{\"o}rnson, J.~Hoydis, L.~Sanguinetti, {\em et~al.}, ``Massive mimo
  networks: Spectral, energy, and hardware efficiency,'' {\em Foundations and
  Trends{\textregistered} in Signal Processing}, vol.~11, no.~3-4,
  pp.~154--655, 2017.

\end{thebibliography}





\end{document}